# A tool for ECG signal analysis using standard and optimized Hermite transform

Zoja Vulaj, Andjela Draganić, Miloš Brajović and Irena Orović
University of Montenegro, Faculty of Electrical Engineering
Džordža Vašingtona bb, 81000 Podgorica, Montenegro
Emails: zojavulaj@gmail.com, andjelad@ac.me, milosb@ac.me, irenao@ac.me

*Abstract*— **The development of a system that would ease the diagnosis of heart diseases would also fasten the work of the cardiologic department in hospitals and facilitate the monitoring of patients with portable devices. This paper presents a tool for ECG signal analysis which is designed in MATLAB. The Hermite transform domain is exploited for the analysis. The proposed transform domain is very convenient for ECG signal analysis and classification. Parts of the ECG signals, i.e. QRS complexes, show shape similarity with the Hermite basis functions, which is one of the reasons for choosing this domain. Also, the information about the signal can be represented using a small set of coefficients in this domain, which makes data transmission and analysis faster. The signal concentration in the Hermite domain and consequently, the number of samples required for signal representation, can additionally be reduced by performing the parametrization of the Hermite transform. For the comparison purpose, the Fourier transform domain is also implemented within the software, in order to compare the signal concentration in two transform domains. The application of the proposed method in clinical practice includes arrhythmia and heart failure detection, as well as other abnormalities of the cardiac rhythm.**

## I. INTRODUCTION

The electrocardiogram (ECG) represents the changes of the electrical activity of the heart over time. By analyzing the ECG signals, the information about the medical condition of the heart is provided. Cardiologists can sometimes determine different abnormalities by observing the electrocardiogram. In some cases, when a high level of accuracy is required, different analysis implemented into an application can provide better results. Due to the specificity of this kind of signals, in everyday application we cannot always rely on the results of the application, but a supervision of cardiologist is required. The development of a system (electrocardiograph) that can work on its own was a great revolution in the world of cardiology [1]. Especially, in the cases of patients real-time monitoring, at home (portable devices) or in hospital, setting an alarm when certain conditions appear could save lives.

ECG signals are periodic signals, because they are composed of a sequence of waves that repeat periodically in time: P wave, then Q, R and S waves (which form the QRS complex) and T wave. Very rarely, a U wave can also be detected. The most characteristic part of an ECG signal is the QRS complex [2]. Particularly, when analyzed and classified, the ECG signals are observed beat-to-beat. The kind of features and the number of signals included into the analysis depend on the type of classification, analysis we are willing to perform, and the results we need.

Signals can be analyzed in different domains: time domain, frequency domain, or the combined time – frequency domain [3]-[6]. There are many different approaches for ECG signal analysis. The choice of domain depends on the requirement of the particular application and information that is necessary.

In this paper, the software for ECG signal representation in different transform domains is proposed. Time domain, the Hermite transform (HT) and the Fourier transform (FT) domain representations are considered [7]. The HT is proven to be a better choice for ECG signal analysis compared to the commonly used FT [8], [9]. Particularly, the ECG signals, i.e., their QRS complexes have a similar shape as the Hermite basis functions. Therefore, it is shown that, by using the HT for the signal analysis, we can retain the signal information using a small amount of coefficients. By introducing two changeable parameters (time shift and scaling factor) the HT can further be advanced to meet the requirements of the signal processing. Therefore, by performing the optimization of the HT, the signal can be additionally sparsified.

The paper is organized as follows: In Section II the theoretical background on the HT is given. The optimization of the HT is given in this section as well. Section III contains the description of the proposed software for the analysis of the ECG signals and its functionality. In Section IV the performance of the proposed software for different types of ECG signals is presented. The concluding remarks can be found in Section V.

## II. THEORETICAL BACKGROUND

### A. *The Hermite transform*

To approximate ECG signals using a reduced number of coefficients, while retaining the accuracy, the Hermite basis functions are considered to be efficient due to the similarity of the ECG and Hermite waveforms. The Hermite functions can be defined in terms of the Hermite polynomials. The *n*-th order Hermite polynomial is given as follows:

$$HP_n(t) = (-1)^n e^{t^2} \cdot \frac{d^n(e^{-t^2})}{dt^n}, \quad (1)$$

while the Hermite functions, in terms of the Hermite polynomial, are defined as:

$$HF_n(t) = e^{-t^2} HP_n(t) / \sqrt{2^n n! \sqrt{\pi}} \ . \quad (2)$$

Generally, in the case of signals that are continuous in time, an infinite number of Hermite functions is required in order to approximate the signal without an error. In real application, the finite number Hermite functions have to be used. As we denoted these functions by $HF_n(t)$, the signal can be defined using the Hermite expansion as follows:

$$\hat{ECG}(t) = \sum_{n=0}^{M-1} Coeff_n HF_n(t) \quad (3)$$

where $M$ is the number of functions used to represent the signal in the HT basis and it can be smaller than the signal length. For ideal signal approximation, the number of Hermite functions should be equal to the signal length $N$. In the discrete domain, the functions are calculated in the roots of the Hermite polynomials [10],[11]. The Hermite coefficients *Coeff* can be calculated using the Hermite polynomials, based on the Gauss-Hermite quadrature method [11]-[16] as:

$$Coeff_n = \frac{1}{\sqrt{2^n n! \sqrt{\pi}}} \sum_{z=1}^{N} \frac{2^{N-1} N! \sqrt{\pi}}{N^2 HP_{N-1}^2(t_z)} \left( ECG(t_z) e^{t_z^2/2} \right) HP_n(t_z) \quad (4)$$

The parameter $t_z$ represent the zeros of the polynomials, while $N$ is the number of samples of the signal $ECG(t)$. The expression for the Hermite coefficients can be significantly simplified if we implement the Hermite functions:

$$Coeff_n \approx \frac{1}{M} \sum_{z=1}^{M} \alpha_{M-1}^n(t_z) f(t_z) , \quad (5)$$

where:

$$\alpha_{M-1}^n(t_z) = HF_n(t_z) / (HF_{M-1}(t_z))^2 . \quad (6)$$

### *B. The parametrization of the Hermite transform*

The Hermite function can be defined by introducing the scaling factor $\delta$, used to match functions to the signal, by stretching or compressing them [8]:

$$HF_n(t,\delta) = e^{-t^2/2\delta^2} HP_n(t/\delta) / \sqrt{\delta 2^n n! \sqrt{\pi}} \ . \quad (7)$$

Beside the scaling factor, the functions can be shifted in time to get the optimal concentration in the HT domain. The rule that we use when determining these parameters in order to get the optimal signal concentration, is based on the $\ell_1$-norm. The value of the scaling factor, is determined experimentally. The signal expansion equation, in this case can be written as:

$$ECG(t) = \sum_{n=0}^{M-1} Coeff_n HF_n(t,\delta) \ . \quad (8)$$

The Gauss-Hermite quadrature equation for calculation of the Hermite coefficients, becomes [11]-[14]:

$$Coeff_n = \frac{1}{N} \sum_{z=1}^{N} \frac{HF(t_z,\delta)}{[HF_{N-1}(t_z,\delta)]^2} ECG(t_z), \ \ n = 0,1,...,N-1 \quad (9)$$

Signals that are being represented by using the HT, are usually sampled at points that are proportional to the Hermite polynomials roots and those points are not uniformly distributed. To obtain the values at the non-uniform points, the interpolation is used [8]:

$$ECG(t_s) \approx \sum_{n=-C}^{C} ECG(n\Delta t) \frac{\sin(\pi(t_s - n\Delta t)/\Delta t)}{\pi(t_s - n\Delta t)/\Delta t}, s = 1,...,N \ , \quad (10)$$

$t_s$ are the sampling points, while $\Delta t$ is the sampling period.

The idea of scaling factor optimization procedure relies on the concentration measure of the Hermite coefficients *Coeff*. Namely, a suitable value of the scaling factor $\delta$ is chosen so that the vector of Hermite coefficients is as sparse as possible. Here, the $\ell_1$-norm of transform coefficients plays the main role. In the case of the HT, this norm can be calculated as follows:

$$\|Coeff\|_1 = \sum_{n=0}^{N-1} |Coeff_n| \quad (11)$$

The optimal value of $\delta$ can now be determined as:

$$\delta = \arg\min_{\delta} \|Coeff\|_1 \quad (12)$$

Using equation (12) we have determined the range of the possible values of the scaling factor. The value that minimizes the concentration measure is optimal. On the other side, the shift parameter $\tau$ used to move the signal for a few samples left or right relative to the zero time instant, can be also considered when searching for an optimal $\delta$. The optimization is done for every possible value of the $\tau$ from the set $[\tau_{min}, \tau_{max}]$, and a measure vector $\mathbf{L}$ is formed. The optimal shift $\tau$ is the one that meets the following equation:

$$\tau = \arg\min_{\tau} \mathbf{L} \ . \quad (13)$$

## III. THE SOFTWARE TOOL FOR THE ECG SIGNAL ANALYSIS

The Graphical User Interface (GUI), i.e., the software tool for the representation of ECG signals and QRS complexes, is designed in Matlab 7. The outlook of the software is displayed in Figure 1. It contains two panels, Healthy and Diseased, for the presentation of two different classes of signals, from two different data sets. The two groups of signals contain nine different test signals (each), and these could be chosen from the dropdown menu.

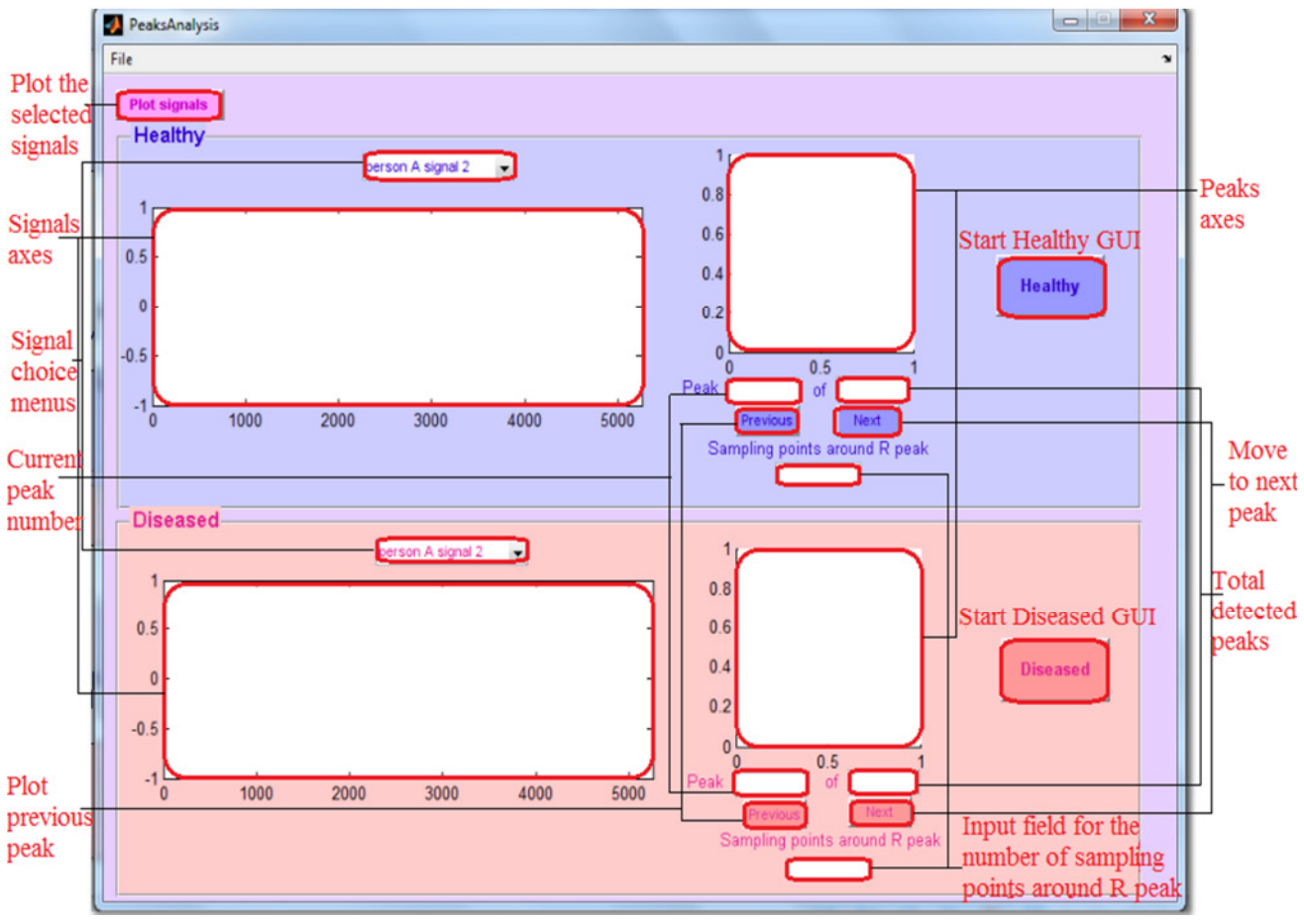

Figure 1. The Virtual Instrument

The chosen test signals can be plot using the button 'Plot signals' – the time domain of the peaks is represented. The peaks axes give us the preview of the extracted peaks for the selected signals. Using the buttons 'Next' and 'Previous', the user can switch among the peaks. On Figure 1, buttons 'Healthy' and 'Diseased' are also marked. When these buttons are clicked, two sub-GUIs are called (Figure 2). Each of them displays the result of the analysis for the selected signal and peak that belongs to the appropriate signal class. The sub-GUI

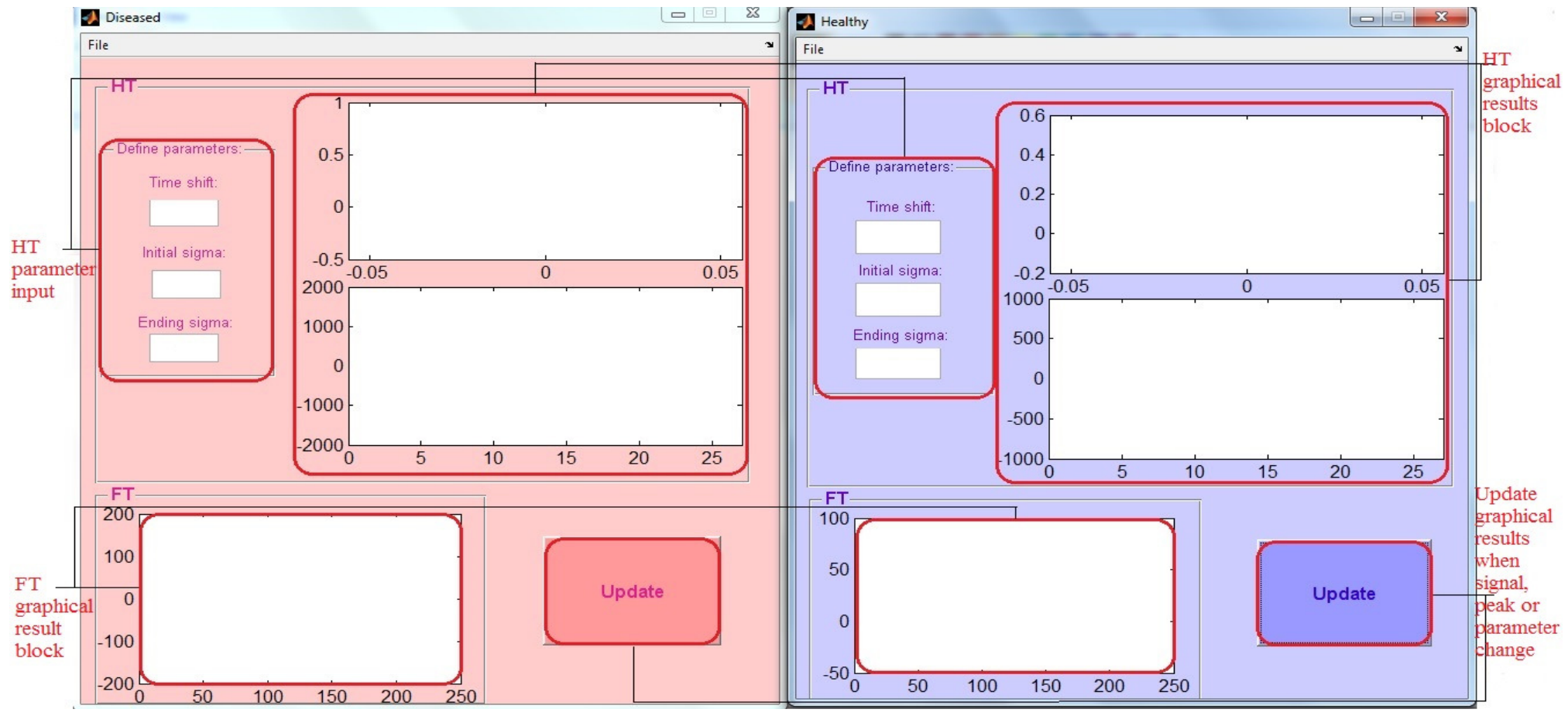

Figure 2. The Healthy and Diseased sub-GUIs

'Healthy', as well as the 'Diseased' one, contain two panels: HT and FT. The panel HT calculates and displays the coefficients of the HT and the coefficients of the optimized HT in the lower axes of the HT panel (Figure 2.). Both types of coefficients, are plot within the same axes, to make the difference between them more noticeable and the comparison of the proposed methods easier. In the upper axes of the same panel on the appropriate sub-GUIs, the representation of the chosen signal part in the Hermite domain, along with the same signal but shifted in time, are shown. In each sub-GUI a sub-panel for the definition of the parameters used in calculating the HT, as well as the optimized HT, is also implemented. After the changes of these parameters, as well as the changes of the signals and desired peaks, the graphics are updated if button 'Update' is clicked. The panel FT contains the axes in which the graphical result of the selected peak is shown when the Fourier transform domain is implemented.

When speaking of the sets of signals used for this analysis, one of the data sets contains the signals and their R peak locations of healthy people, while the other is the data set of the diseased. QRS complexes in ECG signals of healthy and diseased people differ in duration, amplitude and morphology. Depending on these characteristics, the accurate diagnosis of diseases can be established. The extraction of QRS complexes is based, firstly, on the detection of R peaks. For the detection of certain diseases, such as arrhythmia, this information is sufficient. On the other side, when more information is required, it is still based on the R peak since more information, as a rule, means more sampling points around R peak. The number of sampling points, in the presented GUI, can be input by the user in the appropriate field shown on Figure 1. This is implemented in order to compare the changes in the HT and FT when different segments of ECG signals are used.

## IV. EXPERIMENTAL RESULTS

The performance of the main GUI is shown in Figure 3. For the purpose of this analysis, two randomly selected signals are plot and the first peaks are analyzed. After the buttons 'Healthy' and 'Diseased' are clicked, the sub-GUIs are started (Figure 5.). In the panel HT the graphical results of the calculations in the Hermite transform domain are displayed. The coefficients of the HT (marked in blue) and the coefficients of the optimized HT (marked in magenta), are plot within the same axes in order to visually enhance the effects of the HT optimization. The representation of the chosen signal part in the Hermite domain, is also plotted using a blue filled line, along with the preview of the same signal but in magenta and with a certain shift which can be defined by the user. The desired amount of shift, can be input within the same panel. In Figure 4. the changes in the HT and the optimized Hermite transform, for different cases of signal shifting are presented.

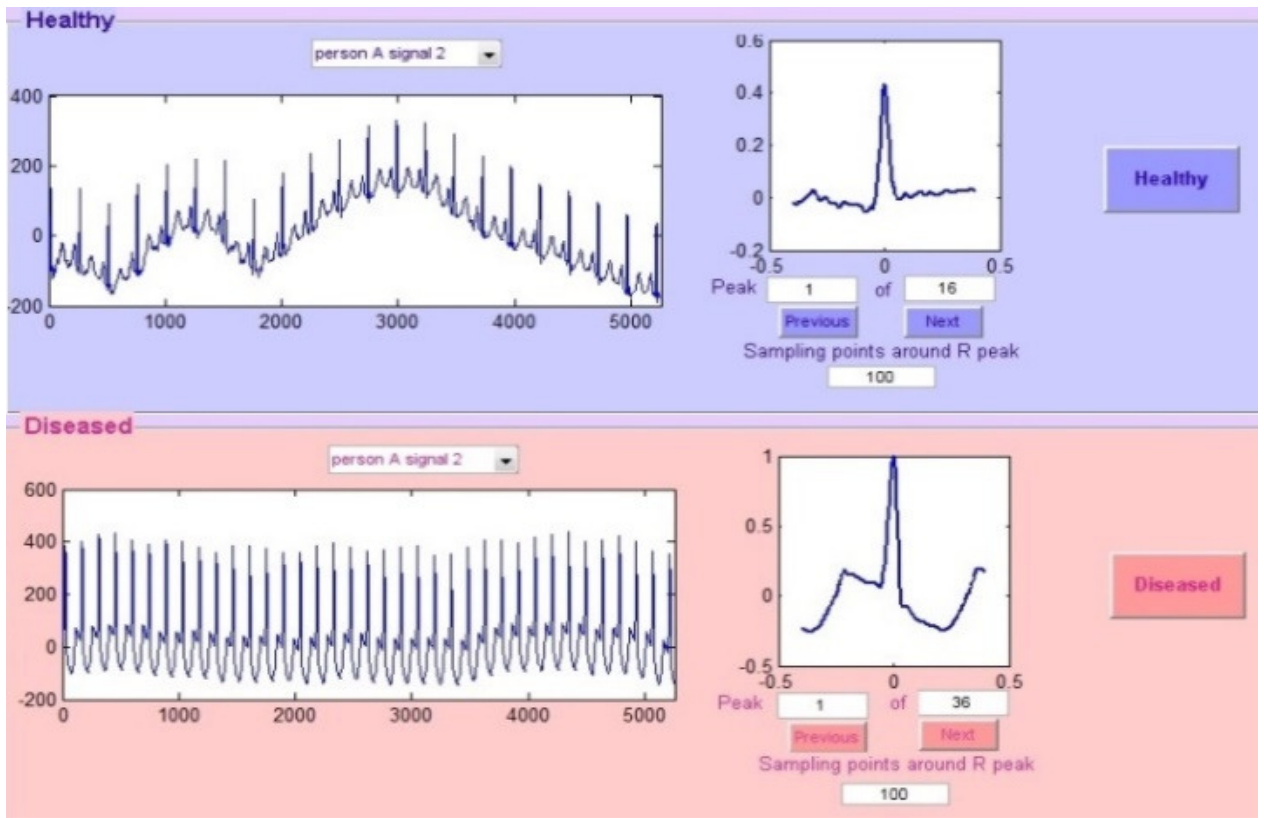

Figure 3. The performance of the instrument

As we can see from the figures, for $\tau$=1 the signals almost overlap in time and the QRS complex, using the parametrized

HT, is represented with the two strongest among all the available coefficients, while retaining the desired accuracy. As we increase the time shift, the values of the coefficients drop significantly but more coefficients need to be used in order to maintain the same accuracy. The effect of different values of the scaling factor $\delta$ on the coefficients of the parametrized HT (the signal chosen from the class of the diseased presented in Figure 3.) are shown on Figure 5. In order to get a better optimization of the HT, two values of the scaling factor are introduced, $\delta_0$ and $\delta_{max}$. These values can be also defined by the user in the fields 'Define parameters', shown in Figure 2. The optimal $\delta$ is then searched in the range:

$$\delta = \delta_0 : 0.1 : \delta_{\max} \quad (14)$$

The step parameter 0.1 is used for the scaling factor optimization [17]. The optimal $\delta$ is the one that enhances the coefficients concentration and meets the sparsity requirements. Note that not every signal will have the same optimal scaling factor even if it belongs to the same signal group, and for each signal there is a certain range of values that $\delta$ can take.

The FT panel, represents the signal in the Fourier transform domain. This domain is also implemented for the comparison with the HT based analysis method. The button 'Update' is used to update the graphics when the user is willing to change the parameters of the analysis.

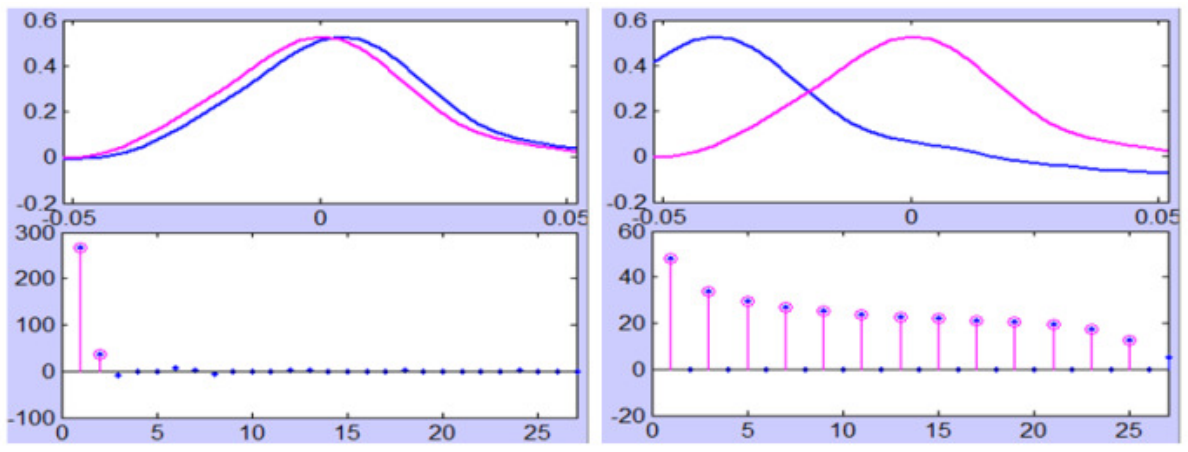

Figure 4. The effect of time shift (the left figure ($\tau$=1) and the right figure ($\tau$=10)).

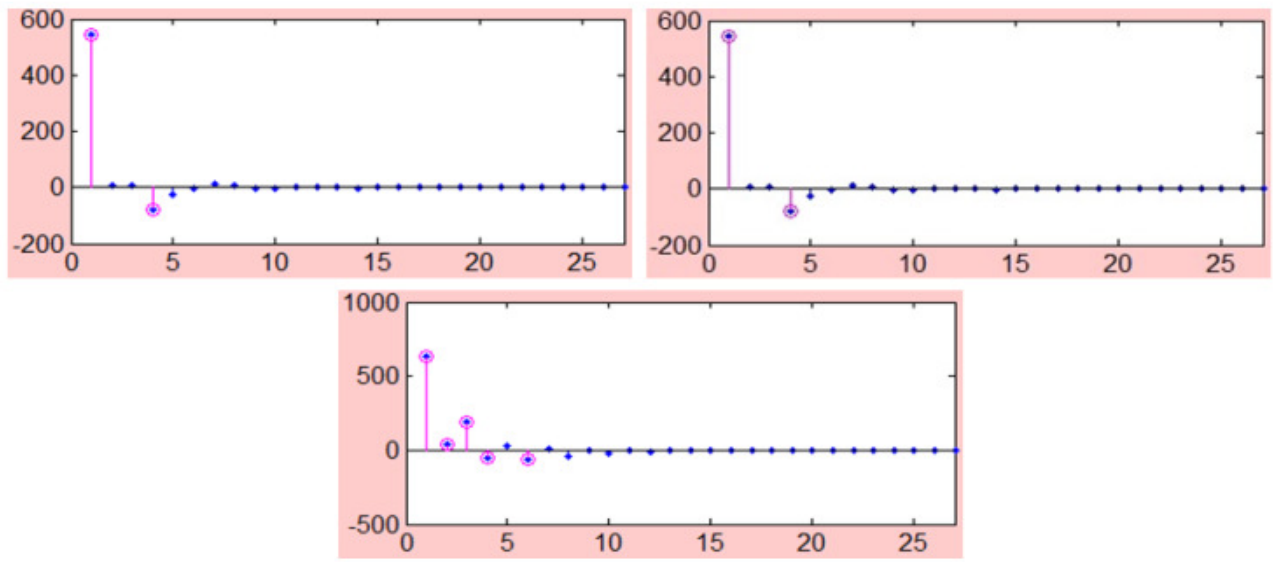

Figure 5. The effect of the scaling factor (first row- left fig. ($\delta_0$=1; $\delta_{max}$=10), right fig. ($\delta_0$=1; $\delta_{max}$=5); second row ($\delta_0$=1; $\delta_{max}$=3).

## V. CONCLUSION

In this paper, the tool implementing the Hermite transform domain is proposed for analysis and representation of the ECG signals, namely their QRS complexes. For the comparison purpose, the Fourier transform is also implemented. The optimization of the Hermite transform is included in the tool, in order to choose suitable shift and scaling parameters. Further research could be oriented to implementing the proposed method for signal classification.

## ACKNOWLEDGMENT

This work is supported by the Montenegrin Ministry of Science, project grant funded by the World Bank loan: CS-ICT "New ICT Compressive sensing based trends applied to: multimedia, biomedicine and communications".